\documentclass[twocolumn,twoside,10pt]{IEEEtran}
\usepackage{setspace}
\usepackage{cite,color}%
\usepackage{amsfonts}%
\usepackage{amsmath,url}%
\usepackage{amssymb}%
\usepackage{graphicx}
\usepackage{balance}
\usepackage{amsmath,amssymb,graphicx,epsfig,url,cite,balance}
\begin{document}
\newcounter{eqncount}
\pagestyle{empty}
\thispagestyle{empty}
\title{On the Statistics of Cognitive Radio Capacity in Shadowing and Fast Fading Environments}
\author{\authorblockA{Muhammad Fainan Hanif~\IEEEauthorrefmark{1}, Peter J.
Smith~\IEEEauthorrefmark{1} and Mansoor
Shafi~\IEEEauthorrefmark{2}}\\
\authorblockA{\IEEEauthorrefmark{1}Department of Electrical and Computer Engineering, University of Canterbury,
Christchurch, New Zealand}
\authorblockA{\IEEEauthorrefmark{2}~Telecom New Zealand, PO Box 293, Wellington, New Zealand\\
Email:mfh21@student.canterbury.ac.nz,~p.smith@elec.canterbury.ac.nz,~mansoor.shafi@telecom.co.nz}}
\maketitle
\thispagestyle{empty}
\pagestyle{empty}
\begin{abstract}
In this paper we consider the capacity of the cognitive radio (CR)
channel in a fading environment under a ``low interference regime".
This capacity depends critically on a power loss parameter,
$\alpha$, which governs how much transmit power the CR dedicates to
relaying the primary message. We derive a simple, accurate
approximation to $\alpha$ which gives considerable insight into
system capacity. We also investigate the effects of system
parameters and propagation environment on $\alpha$ and the CR
capacity. In all cases, the use of the approximation is shown to be
extremely accurate. Finally, we derive the probability that the
``low interference regime" holds and demonstrate that this is the
dominant case, especially in practical CR deployment scenarios.
\end{abstract}
\section{Introduction}
%
The key idea behind the deployment of cognitive radio (CR) is that
greater utilization of spectrum can be achieved if they are allowed
to co-exist with the incumbent licensed primary users (PUs) provided
that they cause minimal interference. The CRs must therefore learn
from the radio environment and adapt their parameters so that they
can co-exist with the primary systems. The CR field has proven to be
a rich source of challenging problems. A large number of papers have
appeared on various aspects of CR, namely spectrum sensing
\cite{Ghasemi1}, fundamental limits of spectrum sharing
\cite{Ghasemi}, information theoretic capacity limits
\cite{devroye,Maric,Jovicic,Jiang} etc.

The 2 user cognitive channel \cite{devroye,Maric,Jovicic,Jiang}
consists of a primary and a secondary user. It is very closely
related to the classic 2 user interference channel, see
\cite{Kramer} and references therein.

The formulation of the CR channel is due to Devroye \emph{et al.}
\cite{devroye}. In this channel, the CR has a non-causal knowledge
of the intended message of the primary and by employing dirty paper
coding \cite{Costa} at the CR transmitter it is able to circumvent
the primary user's interference to its receiver. However, the
interference from the CR to the primary receiver remains and has the
potential to cause a rate loss to the primary.

In recent work, Jovicic and Viswanath \cite{Jovicic} have studied
the fundamental limits of the capacity of the CR channel. They show
that if the CR is able to devote a part of its power to relaying the
primary message, it is possible to compensate for the rate loss to
the primary via this additional relay. They have provided exact
expressions for the PU and CR capacity of a 2 user CR channel when
the CR transmitter sustains a power loss by devoting a fraction,
$\alpha$, of its transmit power to relay the PU message.
Furthermore, they have provided an exact expression for $\alpha$
such that the PU rate remains the same as if there was no CR
interference. It should be stressed here that their system model is
such that at the expense of CR transmit power, the PU device is
always able to maintain a constant data rate. Hence, we focus on CR
rate, $\alpha$ and their statistics. They also assume that the PU
receiver uses a single user decoder. Their result holds for the so
called low interference regime when the received SNR of the CR
transmission is lesser at the primary receiver (i.e., interference
from CR to PU) than at the CR receiver. The authors in \cite{Wu}
also arrived at the same results in their parallel but independent
work.

The Jovicic and Viswanath study is for a static channel, i.e., the
direct and cross link gains are constants. In a system study, these
gains will be random and subject to distance dependent path loss and
shadow fading. Furthermore, the channel gains also experience fast
fading. As the channel gains are random variables, the power loss
parameter, $\alpha$, is also random.

In this paper we focus on the power loss, $\alpha$, the capacity of
the CR channel and the probability  that the ``low interference
regime'' holds. The motivation for this work arises from the fact
that maximum rate schemes for the CR in the low interference regime
\cite{Jovicic} and the achievable rate schemes for the high
interference regime \cite{Maric,Jiang} are very different. Hence, it
is of interest to identify which scenario is the most important. To
attack this question we propose a simple, physically based geometric
model for the CR, PU layout and compute the probability of the low
interference regime. Results are obviously limited to this
particular model but provide some insight into reasonable deployment
scenarios. Since the results show the low interference regime can be
dominant, it is also of interest to characterize CR performance via
the $\alpha$ parameter. In this area we make the following
contributions:
\begin{itemize}
\item Assuming lognormal shadowing, Rayleigh fading and path loss effects we derive the probability  that the ``low interference regime" holds.
\item In the same fading environment we derive an approximation for $\alpha$ and its statistics.
This extremely accurate approximation leads to simple
interpretations of the effect of system parameters on the capacity.
\item Using the statistics of $\alpha$ we investigate the mean rate loss of the
CR and the cumulative distribution function (CDF) of the CR rates.
For both the above we show their dependence on the propagation
parameters.
\item We also show how the mean value of $\alpha$ varies with the CR
transmit power and therefore the CR coverage area.
\end{itemize}
This paper is organized as follows: Section~II describes the system
model. Section~III derives the probability  that the ``low
interference regime" holds and in Section~IV an approximation for
$\alpha$ is developed. Section~V presents analytical and simulation
results and some conclusions are given in Section~VI.
\section{System Model}
\begin{figure}[t]
\centering
\includegraphics[width=0.75\columnwidth]{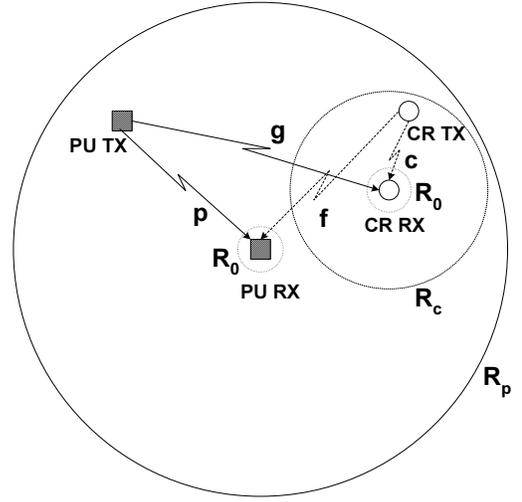}
\caption{System model.} \label{fig_1}
\end{figure}

Consider a PU receiver in the center of a circular region of radius
$R_p$. The PU transmitter is located uniformly in an annulus of
outer radius $R_p$ and inner radius $R_{0}$ centered on the PU
receiver. It is to be noted that we place the PU receiver at the
center only for the sake of mathematical convenience (see Fig.
\ref{fig_1}). The use of the annulus restricts devices from being
too close to the receiver. This matches physical reality and also
avoids problems with the classical inverse power law relationship
between signal strength and distance \cite{Mai1}. In particular,
having a minimum distance, $R_0$, prevents the signal strength from
becoming infinite as the transmitter approaches the receiver.
Similarly, we assume that a CR receiver is uniformly located in the
same annulus. Finally, a CR transmitter is uniformly located in an
annulus centered on the CR receiver. The dimensions of this annulus
are defined by an inner radius, $R_0$, and an outer radius, $R_c$.
Following the work of Jovicic and Viswanath \cite{Jovicic}, the four
channel gains which define the system are denoted $p, g, f, c$. In
this paper, these complex channel gains include shadow fading,
path-loss and Rayleigh fast fading effects. To introduce the
required notation we consider the link from the CR transmitter to
the PU receiver, the CP link. For this link we have:
\begin{equation}\label{linkdef}
|f|^2=\Gamma_{cp}|\tilde{f}|^2,
\end{equation}
where $|\tilde{f}|^2$ is an exponential random variable with unit
mean and $\Gamma_{cp}$ is the link gain. The link gain comprises
shadow fading and distance dependent path loss effects so that,
\begin{equation}\label{signal}
\Gamma_{cp}=A_cL_{cp}r_{cp}^{-\gamma},
\end{equation}
where $A_c$ is a constant, $L_{cp}=10^{\tilde{X}_{cp}/10}$ is
lognormal, $\tilde{X}_{cp}$ is zero mean Gaussian and $r_{cp}$ is
the link distance. The standard deviation which defines the
lognormal is $\sigma$ (dB) and $\gamma$ is the path loss exponent.
For convenience, we also write $L_{cp}=e^{X_{cp}}$ so that
$X_{cp}=\beta \tilde{X}_{cp}$, $\beta=\ln(10)/10$ and
$\sigma_{sf}^2$ is the variance of $X_{cp}$. Hence, for the CP link
we have:
\begin{equation}\label{linkgain}
|f|^2=A_ce^{X_{cp}}r_{cp}^{-\gamma}|\tilde{f}|^2.
\end{equation}
The other three links are defined similarly where $\tilde{p},
\tilde{g}, \tilde{c}$ are standard exponentials, $X_{pp}, X_{pc},
X_{cc},$ are Gaussians with the same parameters as $X_{cp}$ and
$r_{pp}, r_{pc}, r_{cc}$ are link distances. However, for the links
involving PU transmitter we assume a constant $A_p$ in the model of
link gains. The parameters $A_p$ and $A_c$ are  constant and all
links are assumed independent. The remaining parameters required are
the transmit powers of the PU/CR devices, given by $P_p$/$P_c$, and
the noise powers at the PU/CR receivers, given by $N_p$/$N_c$.

For fixed channel coefficients, $p, g, f$ and $c$, Jovicic and
Viswanath \cite{Jovicic} compute the highest rate that the CR can
achieve subject to certain constraints. A key constraint is that the
PU must not suffer any rate degradation due to the CR and this is
achieved by the CR dedicating a portion, $\alpha$, of its transmit
power to relaying the PU message. The parameter, $\alpha$, is
therefore central to determining the CR rate. Furthermore, the
results in \cite{Jovicic} are valid in the ``low interference
regime" defined by $a<1$ where:
\begin{equation}\label{defa}
a=\frac{\sqrt{N_c}\sqrt{\Gamma_{cp}}|\tilde{f}|}{\sqrt{N_p}\sqrt{\Gamma_{cc}}|\tilde{c}|}
=\frac{\sqrt{N_c}e^{X_{cp}/2}r_{cp}^{-\gamma/2}|\tilde{f}|}{\sqrt{N_p}e^{X_{cc}/2}r_{cc}^{-\gamma/2}|\tilde{c}|}.
\end{equation}
In this regime, the highest CR rate is given by
\begin{equation}\label{CRRate}
R_{CR}=\log_2\Bigg(1+\frac{|c|^2(1-\alpha)P_c}{N_c}\Bigg),
\end{equation}
with the power loss parameter, $\alpha$, defined by
\begin{equation}\label{alpha}
\alpha=\frac{|s|^2}{|t|^2}\Bigg[\frac{\sqrt{1+|t|^2(1+|s|^2)}-1}{1+|s|^2}\Bigg]^2,
\end{equation}
where $|s|=\sqrt{P_p}\sqrt{\Gamma_{pp}}|\tilde{p}|N_p^{-1/2}$ and
$|t|=\sqrt{P_c}\sqrt{\Gamma_{cp}}|\tilde{f}|N_p^{-1/2}$. Note that
the definitions of $\alpha$ and $R_c$ are conditional on $a<1$.
Since $a$ is a function of $\tilde{f}$ and $\tilde{c}$ we see that
both $\tilde{f}$ and $\tilde{c}$ are conditional exponentials.
\section{The low interference regime}
The low interference regime is defined by $a<1$, where $a$ is
defined in (\ref{defa}). The probability, $P(a<1)$, depends on the
distribution of $r_{cc}/r_{cp}$. Using standard transformation
theory \cite{pap}, some simple but lengthy calculations show that
the CDF of $r_{cc}/r_{cp}$ is given by (\ref{ratioofdis}).
\begin{figure*}[!t]
\normalsize \setcounter{eqncount}{\value{equation}}
\setcounter{equation}{6}
\begin{equation}\label{ratioofdis}
P\bigg(\frac{r_{cc}}{r_{cp}}<x\bigg) = \left\{ \begin{array}{ll} 0 &
\textrm{$x\leq\frac{R_0}{R_p}$}\\\\
\frac{0.5x^2(R_p^2-R_0^4x^{-4})-R_0^2(R_p^2-R_0^2x^{-2})}{(R_c^2-R_0^2)(R_p^2-R_0^2)}
&
\textrm{$\frac{R_0}{R_p}<x\leq\frac{R_c}{R_p}$}\\\\
\frac{0.5(R_c^4-R_0^4)-R_0^2(R_c^2-R_0^2)+(x^2R_p^2-R_c^2)(R_c^2-R_0^2)}{x^2(R_c^2-R_0^2)(R_p^2-R_0^2)} &
\textrm{$\frac{R_c}{R_p}<x\leq1$}\\\\
1-\frac{0.5R_c^4x^2+0.5R_0^4x^2-R_0^2R_c^2}{(R_c^2-R_0^2)(R_p^2-R_0^2)}
& \textrm{$1<x\leq\frac{R_c}{R_0}$}\\\\
1 & \textrm{$x>\frac{R_c}{R_0}$}
\end{array} \right.
\end{equation}
\setcounter{equation}{\value{eqncount}} \hrulefill
\end{figure*}
\addtocounter{equation}{1}
The CDF in (\ref{ratioofdis}) can be written as:
\begin{equation}\label{simpratioofdis}
P\bigg(\frac{r_{cc}}{r_{cp}}<x\bigg) = c_{i0}x^{-2}+c_{i1}+c_{i2}x^2
\quad \textrm{$i=1,2,3,4,5$}
\end{equation}
where $\Delta=(R_c^2-R_0^2)(R_p^2-R_0^2)$, $c_{10}=0$, $c_{11}=0$,
$c_{12}=0$, $c_{20}=0.5R_0^4/\Delta$, $c_{21}=-R_0^2R_p^2/\Delta$,
$c_{22}=0.5R_p^4/\Delta$, $c_{30}=0.5(R_0^4-R_c^4)/\Delta$,
$c_{31}=R_p^2(R_c^2-R_0^2)/\Delta$, $c_{32}=0$,
$c_{40}=-0.5R_c^4/\Delta$, $c_{41}=1+R_0^2R_c^2/\Delta$,
$c_{42}=-0.5R_0^4/\Delta$, $c_{50}=0$, $c_{51}=1$ and $c_{52}=0$.

\begin{figure}[t]
\centering
\includegraphics[width=0.95\columnwidth]{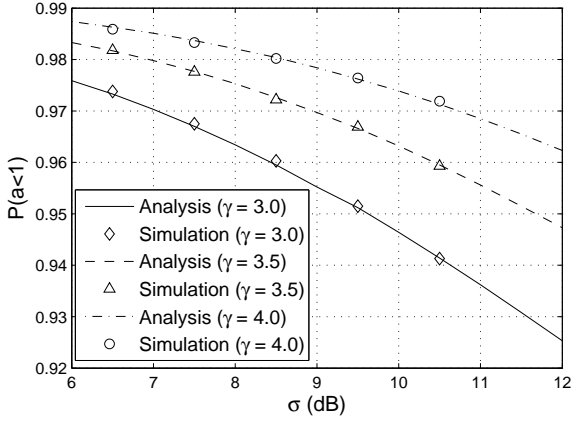}
\caption{Probability of occurrence of the low interference regime as
a function of shadow fading variance, $\sigma$ (dB). The ratio
$R_p/R_c$ is taken as 10.} \label{fig2}
\end{figure}

Now $P(a<1)=P(a^2<1)$ can be written as $P(Y<Ke^XZ^{-\gamma})$ where
$Y={|\tilde{f}|^2}/{|\tilde{c}|^2}$, $K=N_p/N_c$, $X=X_{cc}-X_{cp}$
and $Z=r_{cc}/r_{cp}$. Thus the required probability is:
\setlength{\arraycolsep}{0.0em}
\begin{eqnarray}\label{lowintprob}
P(Y<Ke^XZ^{-\gamma})&{}={}&P(Z<K^{1/\gamma}e^{X/\gamma}Y^{-1/\gamma})\nonumber\\
&{}={}&E[P(Z<K^{1/\gamma}e^{X/\gamma}Y^{-1/\gamma}|X,Y)]\nonumber\\
&{}={}&E[P(Z<W|W)]\nonumber\\
&{}={}&\int_0^\infty P(Z<w)f_W(w)dw,
\end{eqnarray}
\setlength{\arraycolsep}{5pt}
\hspace{-1mm}where $W=K^{1/\gamma}e^{X/\gamma}Y^{-1/\gamma}$ and
$f_W(.)$ is the PDF of $W$. Note that $P(Z<w)$, given in
(\ref{simpratioofdis}), only contains constants and terms involving
$w^{\pm2}$. Hence, we need the following:
\begin{equation}\label{genint}
\int_\theta^\kappa\!\!w^{2m}f_W(w)dw=\int\!\!\int\!
(Ke^xy^{-1})^{2m/\gamma}f_{X,Y}(x,y)dxdy,
\end{equation}
where $m=-1,0,1$ and $f_{X,Y}(.)$ is the joint PDF of $X,Y$. Now,
since $W=K^{1/\gamma}e^{X/\gamma}Y^{-1/\gamma}$, the limits
$\theta\leq w\leq \kappa$ in (\ref{genint}) imply the following
limits for $x$:
\begin{displaymath}
\ln(\theta^{\gamma} K^{-1}y)\leq x\leq \ln(\kappa{^\gamma} K^{-1}y).
\end{displaymath}
Let $\ln(\theta^{\gamma} K^{-1}y)=A$ and $\ln(\kappa{^\gamma}
K^{-1}y)=B$, then noting that $f_{X,Y}(x,y)=f_X(x)f_Y(y)$, the
integral in (\ref{genint}) becomes:
\setlength{\arraycolsep}{0.0em}
\begin{eqnarray}\label{genint2}
\int_\theta^\kappa w^{2m}f_W(w)dw&{}={}&\int_0^\infty
K^{2m/\gamma}y^{-2m/\gamma}f_Y(y)\nonumber\\&&{\times}\:\int_{A}^{B}
e^{2mx/\gamma}f_X(x)dxdy.
\end{eqnarray}
\setlength{\arraycolsep}{5pt}
\hspace{-2mm}Since $X\sim\mathcal{N}(0,2\sigma_{sf}^2)$, the inner
integral in (\ref{genint2}) becomes:
\setlength{\arraycolsep}{0.0em}
\begin{eqnarray}\label{genint3}
\int_{A}^{B}
\!\!\!e^{2mx/\gamma}&{}f_X(x){}&dx=\exp\Bigg(\frac{4m^2\sigma_{sf}^2}{\gamma^2}\Bigg)\nonumber\\&&{\times}\:
\Bigg[\Phi\Bigg(\frac{B-\frac{4m\sigma_{sf}^2}{\gamma}}{\sqrt{2}\sigma_{sf}}\Bigg)-
\Phi\Bigg(\frac{A-\frac{4m\sigma_{sf}^2}{\gamma}}{\sqrt{2}\sigma_{sf}}\Bigg)\Bigg],\nonumber\\
\end{eqnarray}
\setlength{\arraycolsep}{5pt}
\hspace{-1.6mm}where $\Phi$ is the CDF of a standard Gaussian. Since
$f_Y(y)$ is the density function of the ratio of two standard
exponentials, it is given by \cite{Ghasemi}:
\begin{equation}\label{ratioexp}
f_Y(y)=\frac{1}{(1+y)^2}, \qquad y\geq 0
\end{equation}
Using (\ref{genint3}) and (\ref{ratioexp}), the total general
integral in (\ref{genint}) becomes:
\setlength{\arraycolsep}{0.0em}
\begin{eqnarray}\label{genintfinal}
\int_\theta^\kappa\!\!w^{2m}f_W(w)dw&{}={}&\int_0^\infty
\!\!K^{2m/\gamma}y^{-2m/\gamma}(1+y)^{-2}\exp\Bigg(\frac{4m^2\sigma_{sf}^2}{\gamma^2}\Bigg)\nonumber\\&&{\times}\:
\Bigg[\Phi\Bigg(\frac{B-\frac{4m\sigma_{sf}^2}{\gamma}}{\sqrt{2}\sigma_{sf}}\Bigg)-
\Phi\Bigg(\frac{A-\frac{4m\sigma_{sf}^2}{\gamma}}{\sqrt{2}\sigma_{sf}}\Bigg)\Bigg]dy\nonumber\\
&{}\triangleq{}&I(m,\theta,\kappa).
\end{eqnarray}
\setlength{\arraycolsep}{5pt}
Substituting (\ref{simpratioofdis}) and (\ref{genintfinal}) in
(\ref{lowintprob}) gives $P(a<1)$ as:
\setlength{\arraycolsep}{0.0em}
\begin{eqnarray}\label{lowintfinal}
P(a<1)&{}={}&P(Y<Ke^XZ^{-\gamma})\nonumber\\
&{}={}&\sum_{i=2}^5c_{i0}I(-1,\theta_i,\kappa_i)+c_{i1}I(0,\theta_i,\kappa_i)+c_{i2}I(1,\theta_i,\kappa_i)\nonumber\\
&{}={}&\sum_{i=2}^5\sum_{j=0}^2c_{ij}I(j-1,\theta_i,\kappa_i).
\end{eqnarray}
\setlength{\arraycolsep}{5pt}
\hspace{-1.0mm}Finally, it can be seen from the limits given in
(\ref{ratioofdis}) that $\kappa_i=\theta_{i+1}$. Hence, the final
expression for the probability of occurrence of the low interference
regime is:
\setlength{\arraycolsep}{0.0em}
\begin{eqnarray}\label{lowintfinal1}
P(a<1)&{}={}&\sum_{i=2}^5\sum_{j=0}^2c_{ij}I(j-1,\theta_i,\theta_{i+1}),
\end{eqnarray}
\setlength{\arraycolsep}{5pt}
\hspace{-1.5mm}where the $c_{ij}$ were defined after
(\ref{simpratioofdis}), $I(j-1,\theta_i,\theta_{i+1})$ is given in
(\ref{genintfinal}), $\theta_2=R_0/R_p$, $\theta_3=R_c/R_p$,
$\theta_4=1$, $\theta_5=R_c/R_0$ and $\theta_6=\infty$. Hence,
$P(a<1)$ can be derived in terms of a single numerical integral. For
numerical convenience, (\ref{genintfinal}) is rewritten using the
substitution $v=y(y+1)^{-1}$ so that a finite range integral over
$0<v<1$ is used for numerical results:
\setlength{\arraycolsep}{0.0em}
\begin{eqnarray}\label{genintfinalsim}
\int_\theta^\kappa\!\!\!w^{2m}&{}f_W(w)dw{}&\:=\int_0^1
\!\!\!K^{2m/\gamma}\Big(\frac{v}{1-v}\Big)^{-2m/\gamma}\exp\Bigg(\frac{4m^2\sigma_{sf}^2}{\gamma^2}\Bigg)\nonumber\\&&{\times}\:
\Bigg[\Phi\Bigg(\frac{B-\frac{4m\sigma_{sf}^2}{\gamma}}{\sqrt{2}\sigma_{sf}}\Bigg)-
\Phi\Bigg(\frac{A-\frac{4m\sigma_{sf}^2}{\gamma}}{\sqrt{2}\sigma_{sf}}\Bigg)\Bigg]dv\nonumber\\
&{}\triangleq{}&I(m,\theta,\kappa),
\end{eqnarray}
\setlength{\arraycolsep}{5pt}
\hspace{-1.68mm}where $\ln(\theta^{\gamma} K^{-1}\frac{v}{1-v})=A$
and $\ln(\kappa{^\gamma} K^{-1}\frac{v}{1-v})=B$. Further
simplification of $(\ref{genintfinal})$ appears difficult but the
result in (\ref{genintfinalsim}) is stable and rapid to compute.
A comparison of simulated and analytical results is shown in Fig.
\ref{fig2}. It can the seen that the analytical formula given in
(\ref{genintfinalsim}) perfectly matches the simulation results.
\section{An Approximation For The Power Loss Parameter}
In this section we focus on the power loss parameter, $\alpha$,
which governs how much of the transmit power the CR dedicates to
relaying the primary message. The exact distribution of $\alpha$
appears to be rather complicated, even for fixed link gains (fixed
values of $\Gamma_{cp},\Gamma_{pc},\Gamma_{pp}$ and $\Gamma_{cc}$).
Hence, we consider an extremely simple approximation based on the
idea that $|s|\times|t|$ is usually small and $|s|\times|t|>>|t|$.
This approximation is motivated by the fact that the CP link is
usually very weak compared to the PP link. This is because the CRs
will employ much lower transmit powers than the PU. With this
assumption it follows that $|t|^2(1+|s|^2)$ is small and we have:
\setlength{\arraycolsep}{0.0em}
\begin{eqnarray}\label{alphasimplify}
\sqrt{\alpha}&{}={}&\frac{|s|}{|t|}\Bigg[\frac{\big(1+|t|^2(1+|s|^2)\big)^{1/2}-1}{1+|s|^2}\Bigg]\nonumber\\
&{}\approx{}&\frac{|s|}{|t|}\Bigg[\frac{1/2|t|^2(1+|s|^2)}{1+|s|^2}\Bigg]\nonumber\\
&{}={}&\frac{|s||t|}{2}\nonumber\\
&{}={}&\sqrt{\alpha_{approx}}.
\end{eqnarray}
\setlength{\arraycolsep}{5pt}
\noindent Expanding $\alpha_{approx}$ we have:
\begin{equation}\label{alphaapprox}
\alpha_{approx}= \frac{A_pA_cP_p P_c}{4
N_p^2}e^{(X_{pp}+X_{cp})}r_{pp}^{-\gamma}r_{cp}^{-\gamma}|{\tilde{p}}|^2|{\tilde{f}}|^2.
\end{equation}

\begin{figure}[t]
\centering
\includegraphics[width=0.95\columnwidth]{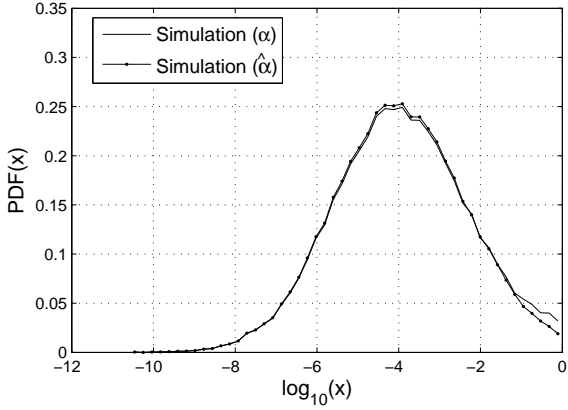}
\caption{PDFs of $\log_{10}(\alpha)$ and its approximation
$\log_{10}(\hat{\alpha})$.} \label{fig3}
\end{figure}

This approximation is very effective for low values of
$\alpha_{approx}$, but is poor for larger values since
$\alpha_{approx}$ is unbounded whereas $0<\alpha<1$. To improve the
approximation, we use the conditional distribution of
$\alpha_{approx}$ given that $\alpha_{approx}<1$. This conditional
variable is denoted, ${\hat{\alpha}}$. The exact distribution of
${\hat{\alpha}}$ is difficult for variable link gains. However, the
approximation has a simple representation which leads to
considerable insight into the power loss and how it relates to
system parameters. For example $\alpha_{approx}$ is proportional to
$|s|^2|t|^2$ so that high power loss may be caused by high values of
$|s|$ or $|t|$ or moderate values of both. Now $|s|$ and $|t|$
relate to the PP and CP links respectively. Hence the CR is forced
to use high power relaying the PU message when the CP link is
strong. This is obvious as the relay action needs to make up for the
strong interference caused by the CR. The second scenario is that
the CR has high $\alpha$ when the PP link is strong. This is less
obvious, but here the PU rate is high and a substantial relaying
effort is required to counteract the efforts of interference on a
high rate link. This is discussed further in Section~V. It is worth
noting that the condition $|s||t|>>|t|$ holds good only for some
specific values of channel parameters. Hence, although it is
motivated by a sensible physical scenario, it certainly needs
checking. Results in Figs.~\ref{fig3}, \ref{fig6} and \ref{fig4}
show that it works very well.

\vspace{-0.20mm} For fixed link gains, the distribution of
${\hat{\alpha}}$ is:
\setlength{\arraycolsep}{0.0em}
\begin{eqnarray}\label{begin}
P(\alpha_{approx}<x|\alpha_{approx}<1)&{}={}&P(\hat{\alpha}<x)\nonumber\\
&{}={}&\frac{P(\alpha_{approx}<x)}{P(\alpha_{approx}<1)}.
\end{eqnarray}
\setlength{\arraycolsep}{5pt}
\noindent Thus, to compute the distribution function of
$\hat{\alpha}$ we need to determine $P(\alpha_{approx}<x)$ which can
be written as
\begin{equation}\label{firststep}
P(\alpha_{approx}<x)=P(|s|^2|t|^2<4x).
\end{equation}
In the analytical approximation below we assume that $|s|^2$ and
$|t|^2$ are exponential, i.e., we ignore the conditioning on $a<1$.
The conditioning can be handled exactly but results suggest that a
simple exponential approximation is satisfactory. Let
$E(|s|^2)=\mu_s$, $E(|t|^2)=\mu_t$ with $\mu_s=P_p\Gamma_{pp}/N_p$
and $\mu_t=P_c\Gamma_{cp}/N_p$. Further, suppose that $U$ and $V$
represent i.i.d.  standard exponentials, then we have
\setlength{\arraycolsep}{0.0em}
\begin{eqnarray}\label{approxcdf}
P(\alpha_{approx}<x)&{}={}&P\bigg(UV<\frac{4x}{\mu_s\mu_t}\bigg)\nonumber\\
&{}={}&E_V\bigg(P\bigg(U<\frac{4x}{V\mu_s\mu_t}\bigg)\bigg)\nonumber\\
&{}={}&E_V\bigg(1-\exp\bigg(\frac{-4x}{V\mu_s\mu_t}\bigg)\bigg)\nonumber\\
&{}={}&1-\int_0^\infty\exp\bigg(\frac{-4x}{v\mu_s\mu_t}-v\bigg)dv\nonumber\\
&{}={}&1-\sqrt{\frac{16x}{\mu_s\mu_t}}K_1\bigg(\sqrt{\frac{16x}{\mu_s\mu_t}}\bigg),
\end{eqnarray}
\setlength{\arraycolsep}{5pt}
\hspace{-1.5mm}where $K_1(.)$ represents the modified Bessel
function of the second kind and the integral in (\ref{approxcdf})
can be found in \cite{int}. Using the expression given in
(\ref{approxcdf}), the CDF of $\hat{\alpha}$ follows from
(\ref{begin}). Note that the CDF of $R_c$ can easily be obtained in
the form of a single numerical integral for fixed powers.
\section{Results}
\begin{figure}[t]
\centering
\includegraphics[width=0.95\columnwidth]{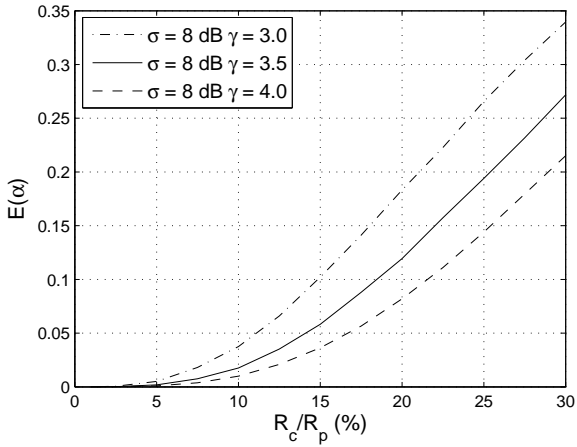}
\caption{Mean value of the power loss parameter, $\alpha$, as a
function of the ratio $\frac{R_c}{R_p}$.} \label{fig5}
\end{figure}
\begin{figure}[t]
\vspace{0mm}\centering
\includegraphics[width=0.95\columnwidth]{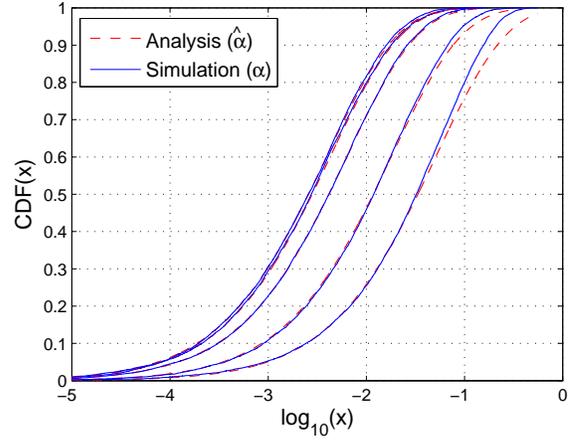}
\caption{Comparison of the exact and analytical CDFs of the power
loss factor on a logarithmic scale for fixed link gains. Results are
shown for 5 drops.} \label{fig6}
\end{figure}
\begin{figure}[t]
\vspace{-3.3mm}\centering
\includegraphics[width=0.97\columnwidth,height=63.5mm]{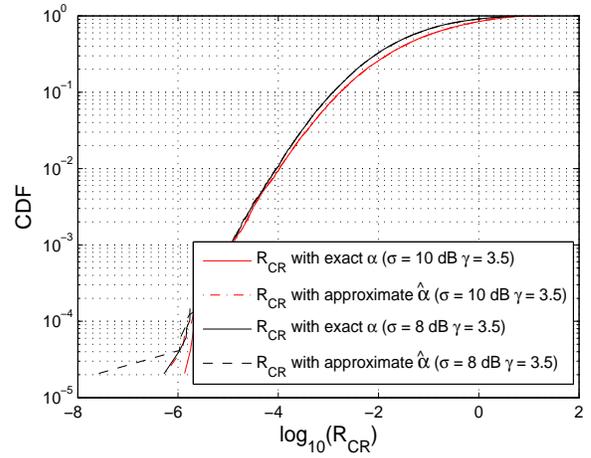}
\caption{CDF of the CR rates with the exact $\alpha$ and the
approximate $\hat{\alpha}$.} \label{fig4}
\end{figure}
\begin{figure}[t]
\centering
\includegraphics[width=0.95\columnwidth]{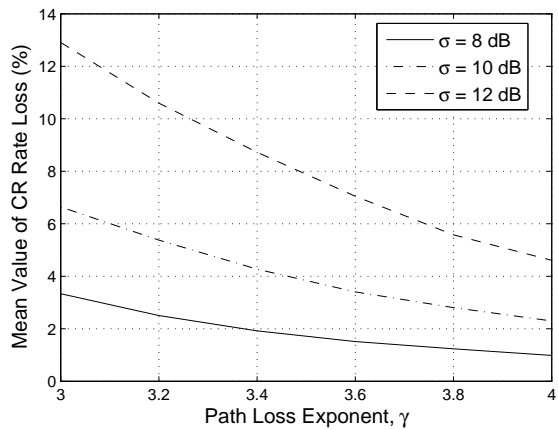}
\caption{Mean value of the CR rate loss as a function of $\gamma$.}
\label{fig7}
\end{figure}
\begin{figure}[t]
\centering
\includegraphics[width=0.99\columnwidth]{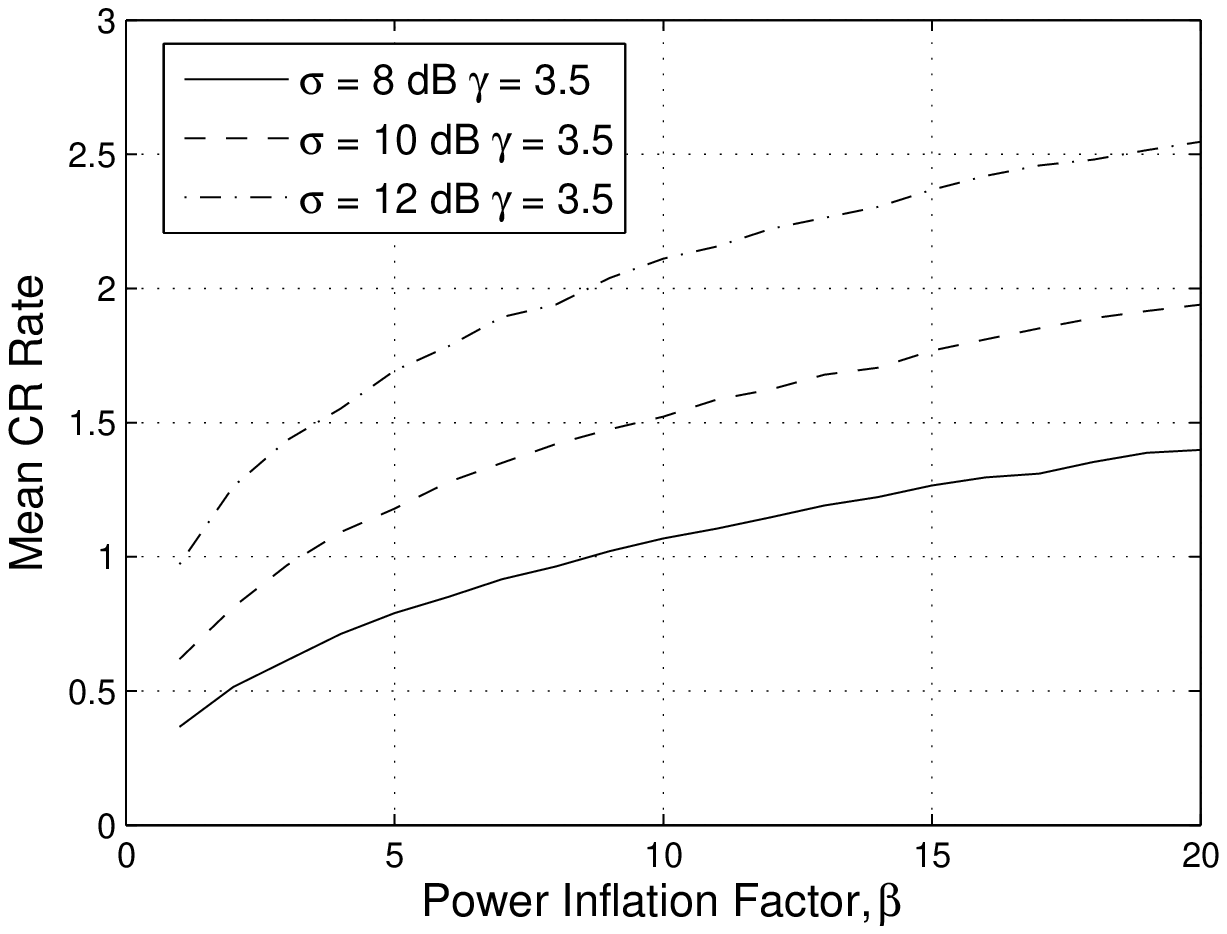}
\caption{Variation of the mean CR rate with the power inflation
factor, $\beta$.} \label{fig8}
\end{figure}
In the results section, the default parameters are $\sigma=8$ dB,
$\gamma=3.5$, $R_0=1$, $R_c=100$ m, $R_p=1000$ m and
$N_p=N_c=P_p=P_c=1$. The parameter $A_p$ is determined by ensuring
that the link PP has an SNR $\geq5$ dB 95\% of the time in the
absence of any interference. Similarly, assuming that both PU and CR
devices have same threshold power at their cell edges, the constant
$A_c=A_p(R_p/R_c)^{-\gamma}$. Unless otherwise stated these
parameters are used in the following.
\subsection{Low interference regime}
In Fig.~\ref{fig2} we show that the low interference
regime, $a<1$, is the dominant scenario. For typical values of
$\gamma\in[3,4]$ and $\sigma\in[6,12]$ dB we find that $P(a<1)$ is
usually well over 90\%. Figure~\ref{fig2} also verifies the
analytical result in (\ref{lowintfinal}).

The relationship between $P(a<1)$ and the system parameters is
easily seen from (\ref{defa}) which contains the term
$\big(r_{cc}/r_{cp}\big)^{\gamma/2}\exp\big((X_{cc}-X_{cp})/2\big)$.
When $R_c<<R_p$, this term decreases dramatically as $\gamma$
increases and as $\sigma$ increases the term increases. Also, as
$R_c$ increases $r_{cc}/r_{cp}$ tends to increase which in turn
increases $P(a<1)$. When $R_c\approx R_p$ the low and high
interference scenarios occur with similar frequency. This may be a
relevant system consideration if CRs were to be introduced in
cellular bands where the cellular hot spots, indoor micro-cells and
CRs will have roughly the same coverage radius. Note that $a$ is
independent of the transmit power, $P_c$. These conclusions are all
verified by simulations which are omitted for reasons of space.
\subsection{Statistics of the power loss parameter, $\alpha$}
Figures~3-5 all focus on the properties of $\alpha$. Figure
\ref{fig3} shows that the probability density function (PDF) of
$\alpha$ is extremely well approximated by the PDF of
$\hat{\alpha}$. In Fig.~\ref{fig5} we see that $E(\alpha)$ increases
with increasing values of $R_c/R_p$ and decreasing values of
$\gamma$. This can be seen from (\ref{alphaapprox}) where
$\alpha_{approx}$ contains a $(r_{pp}r_{cp})^{-\gamma}$ term which
increases as $\gamma$ decreases. The increase of $E(\alpha)$ with
$R_c$ follows from the corresponding increase in $P_c$ to cater for
larger $R_c$ values. In Fig.~\ref{fig5} we have limited $R_c/R_p$ to
a maximum of $30\%$ as beyond this value the high interference
regime is also present with a non-negligible probability. In
Fig.~\ref{fig6} we see the analytical CDF in (\ref{approxcdf})
verified by simulations for five different scenarios of fixed link
gains (simply the first five simulated values of $\Gamma_{pp}$ and
$\Gamma_{cp}$). Note that in the different curves each correspond to
a random drop of the PU and CR transmitters. This fixes the distance
and shadow fading terms in the link gains in (\ref{signal}), thereby
the remaining variation in (\ref{linkdef}) is only Rayleigh. By
computing a large number of such CDFs and averaging them over the
link gains a single CDF can be constructed. This approach can be
used to find the PDF of $\hat{\alpha}$ as shown in Fig.~\ref{fig3}.
Note that the curves in Fig.~\ref{fig6} do not match exactly since
the analysis is for $\hat{\alpha}$ and the simulation is for
$\alpha$.
\subsection{CR rates}
Figures~6-8 focus on the CR rate $R_{CR}$. Figure~\ref{fig4}
demonstrates that the use of $\hat{\alpha}$ is not only accurate for
$\alpha$ but also leads to excellent agreement for the CR rate,
$R_{CR}$. This agreement holds over the whole range and for all
typical parameter values. Figure \ref{fig7} shows the \% loss given
by $[R_{CR}(\alpha=0)-R_{CR}(\alpha)]/[R_{CR}(\alpha=0)]\%$. The
loss decreases as $\gamma$ increases, as discussed above, and
increases with $\sigma$. From (\ref{alphaapprox}) it is clear that
increasing $\sigma$ lends to larger values of $\exp(X_{pp}+X_{cp})$
which in turn increases $\alpha$ and the rate loss. Note that the
rate loss is minor for $\sigma\in[8-10]$ dB with $R_c=R_p/10$. In a
companion paper \cite{ICC}, we show that the interference to the PU
increases with $\sigma$ and decreases with $\gamma$. These results
reinforce this observation, i.e., when the PU suffers more
interference ($\sigma$ is larger) the CR has to devote a higher part
of its power to the PU. Consequently the percentage rate loss is
higher.

Finally, in Fig.~\ref{fig8} we investigate the gains available to
the CR through increasing transmit power. The original transmit
power, $P_c$, is scaled by $\beta$ and the mean CR rate is simulated
over a range of $\beta$ values. Due to the relaying performed by the
CR, the PU rate is unaffected by the CR for any values of $\beta$
and so the CR is able to boost its own rate with higher transmit
power. Clearly the increased value of $\alpha$ for higher values of
$\beta$ is outweighed by the larger $P_c$ value and so the CR does
achieve an overall rate gain. In a very coarse way these results
suggest that multiple CRs may be able to co-exist with the PU since
the increased interference power might be due to several CRs and the
rate gain might be spread over several CRs. Of course, this
conclusion is speculative as the analysis is only valid for a single
CR.
\section{Conclusion}
In this paper we derive the probability that the ``low interference
regime'' holds and demonstrate the conditions under which this is
the dominant scenario. We show that the probability of the low
interference regime is significantly influenced by the system
geometry. When the CR coverage radius is small relative to the PU
radius, the low interference regime is dominant. When the CR
coverage radius approaches a value similar to the PU coverage
radius, the low and high interference regimes both occur with
roughly equal probability. In addition we have derived a simple,
accurate approximation to $\alpha$ which gives considerable insight
into the system capacity. The $\alpha$ approximation shows that CR
rates are reduced by large CR coverage zones, small values of
$\gamma$ and large values of $\sigma$. Finally, we have shown that
the CR can increase its own rate with higher transmit powers,
although the relationship is only slowly increasing as expected.
\balance
\bibliographystyle{IEEEtran}
\bibliography{IEEEabrv,jovicic}
\end{document}